\documentclass{iaus}
\usepackage{graphicx}

\title[``Multiple Multipole'' shear measurement]
{Shapelets\\``multiple multipole'' shear measurement}

\author[Massey et al.]%
{Richard Massey$^1$, Alexandre Refregier$^2$ \and David Bacon$^3$}

\affiliation{
$^1$ California Institute of Technology, 1200 E. California Blvd., Pasadena, CA 91125, U.S.A. \break email: rjm@astro.caltech.edu\\[\affilskip]
$^2$ Service d'Astrophysique, B\^{a}t. 709, CEA Saclay, F-9119 Gif sur Yvette,
France \break $^3$ Institute for Astronomy, Blackford Hill, Edinburgh EH9 3HJ,
U.K.} 

\pubyear{2004}
\volume{225}
\pagerange{1--8}
\date{?? and in revised form ??}
\jname{Impact of Gravitational Lensing on Cosmology}
\editors{Mellier, Y. \& Meylan,G. eds.}

\begin{document}

\maketitle

\begin{abstract} The measurement of weak gravitational lensing is currently
limited to a precision of $\sim$10\% by instabilities in galaxy shape
measurement techniques and uncertainties in their calibration. The potential of
large, on-going and future cosmic shear surveys will only be realised with the
development of more accurate image analysis methods. We present a description
of several possible shear measurement methods using the linear ``shapelets''
decomposition. Shapelets provides a complete reconstruction of any galaxy
image, including higher-order shape moments that can be used to generalise the
KSB method to arbitrary order. Many independent shear estimators can then be
formed for each object, using linear combinations of shapelet coefficients.
These estimators can be treated separately, to improve their overall
calibration; or combined in more sophisticated ways, to eliminate various
instabilities and a calibration bias. We apply several methods to simulated
astronomical images containing a known input shear, and demonstrate the
dramatic improvement in shear recovery using shapelets. A complete IDL software
package to perform image analysis and manipulation in shapelet space can be
downloaded from {\tt www.astro.caltech.edu/$\sim$rjm/shapelets/}.
\end{abstract}

\firstsection 
\firstsection\section{Requirements for a shear estimator}
\label{intro}

Mass fluctuations along the line of sight to a distant galaxy distort its
apparent shape via weak gravitational lensing. If we can measure the ``shear''
field $\gamma$ from the observed shapes of galaxies, we can map out the
intervening mass distribution. But how should the galaxies' shapes be measured?

A monochromatic image of the sky is simply a two-dimensional function of
surface brightness, in which the galaxies are isolated peaks. We would like to
form local shear estimators $\hat{\gamma}$ from some combination of the pixel
values around each peak. The estimators are merely required to trace the true
shear signal when averaged over a galaxy population:
$\langle\hat{\gamma}\rangle=\gamma$. Individual estimators will inevitably be
noisy, because of galaxies' wide range of intrinsic ellipticities and
morphologies. Furthermore, we are primarily interested in distant (and
therefore faint) galaxies. Additional biases from observational noise can
therefore be limited by forcing $\hat{\gamma}$ to be a linear (or only mildly
non-linear) combination of the pixel values.

The standard shear measurement method applied to most current weak lensing data
was invented by \cite[Kaiser, Squires \& Broadhurst (1995; KSB)]{ksb}. KSB
provides a formalism to correct for smearing by a Point-Spread Function (PSF),
and to form a shear estimator $\hat{\gamma}\equiv e/P^\gamma$. It uses a
galaxy's Gaussian-weighted quadrupole ellipticity $e$, because the unweighted
ellipticity does not converge in the presence of observational noise.
Unfortunately, the weight function complicates PSF correction, and there is no
obvious choice for its scale size. It is important to note that such an 
ellipticity by itself would \textit{not} be a valid shear estimator. It does
not respond linearly with shear; nor is it expected to, and this is a separate
issue from the 0.85 calibration factor of \cite{baconsims}. The necessary
``shear susceptibility'' factor, $P^\gamma$, is calculated from the object's
higher-order moments. Heymans (2004; poster at this conference) finds that most
practical problems with the KSB method arise during the measurement of
$P^\gamma$. It can be noisy (the distribution of $\hat\gamma$ then obtains
large wings that need to be artificially truncated for
$\langle\hat\gamma\rangle$ to converge); it is a tensor (for which division is
mathematically ill-defined, or inversion numerically unstable); it assumes the
object is intrinsically circular (to eliminate the off-diagonal terms in the
tensor); and it needs to be measured from an image \textit{before} the shear is
applied. The last two problems can never be solved for an individual galaxy
because it is impossible to observe the pre-shear sky. They are circumvented by
fitting $P^\gamma$ from many galaxies, as a function of their size, magnitude
(\textit{and ellipticity!}) in a sufficiently wide area to contain no coherent
shear signal. However, these steps restrict KSB to a non-local combination of
galaxy shapes in a large population ensemble, introduce the problem of ``Kaiser
flow'' (Kaiser 2000), and also tend to introduce biases of around ten percent.

The potential of modern, high resolution imaging surveys to accurately measure
shear and reconstruct the mass distribution of the universe is now limited by
the precision of KSB. Several efforts are under way to invent new shear
estimators and shear measurement methods to take advantage of such data
(\cite[Bridle et al. 2004]{im2shape}, \cite[Bernstein \& Jarvis 2002]{bj02},
\cite[Goldberg \& Bacon 2004]{octo}, \cite[Refregier \& Bacon
2003]{shapelets2}).

\firstsection \section{Shapelets}

Among the most promising candidates to supercede KSB are shapelets-based
analysis methods (\cite[Refregier 2003]{shapelets1}, \cite[Massey \& Refregier
2004]{shapelets3}). Indeed, the shapelets formalism is a logical extension of
KSB, introducing higher order terms that can be used to not only  increase the
accuracy of the older method, but also to remove its various biases. Shapelets
has already proved useful for image compression and simulation (\cite[Massey et
al. 2004]{shims}) and the quantitative parameterisation of galaxy morphologies
(\cite[Kelly \& McKay 2004]{sdssa}). It seems reasonable that if it can
parameterise the unlensed shapes of galaxies, it should also be able to measure
small perturbations in these shapes.

\begin{figure} \begin{center}
\includegraphics[height=90mm]{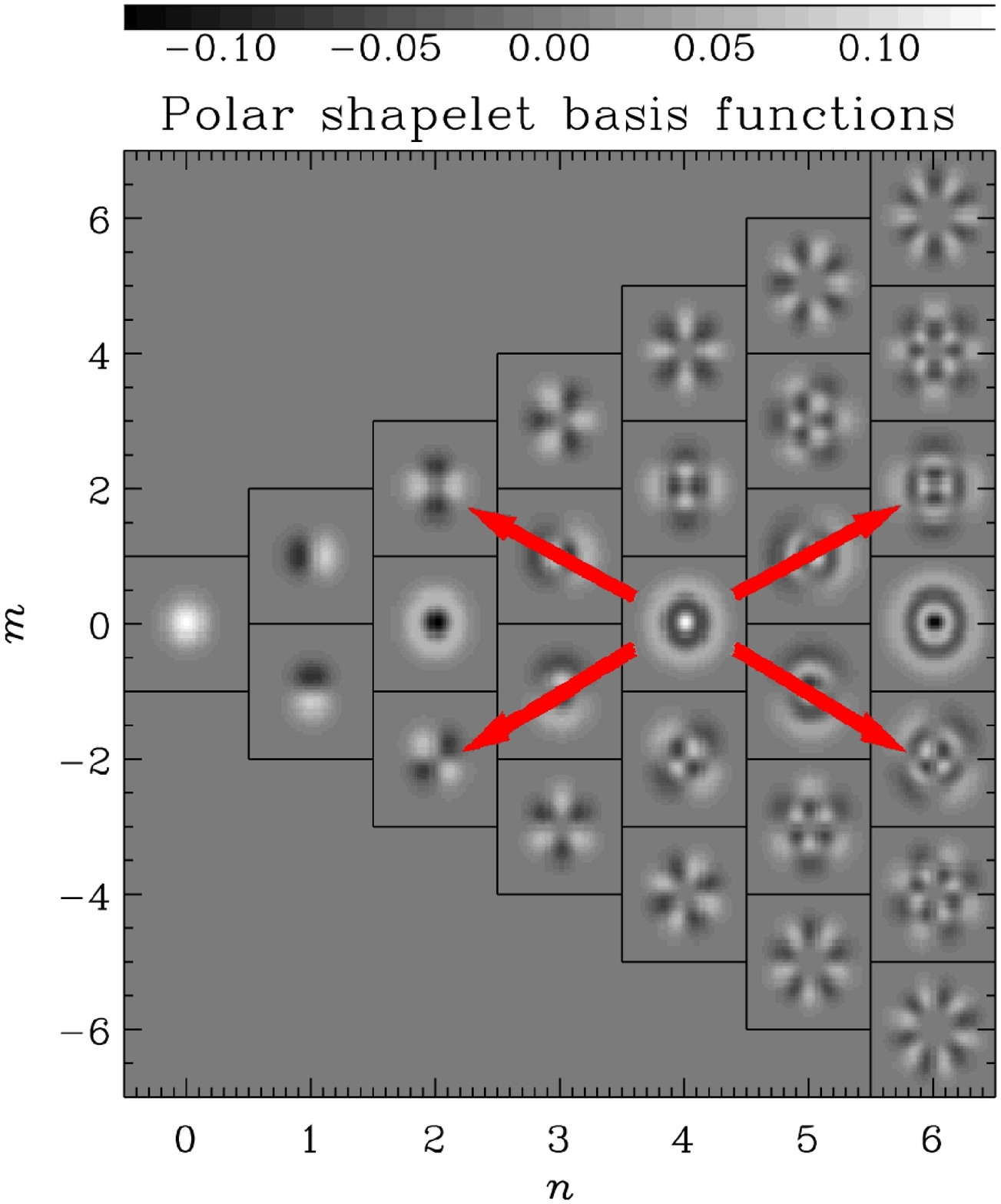}~~~~
\includegraphics[height=90mm]{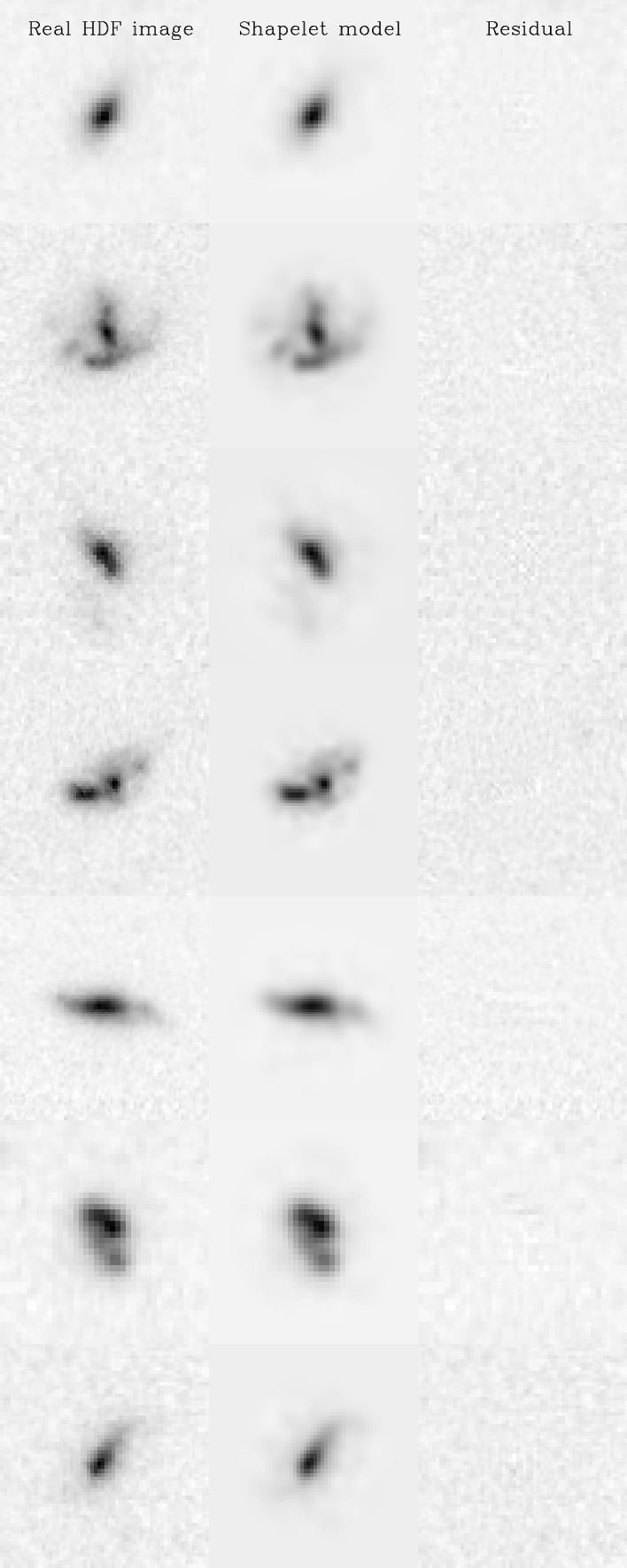} \end{center}
\caption{\textit{Left panel:} Polar shapelet basis functions. These are complex
functions: the real parts are shown in the top half ($m\geq0$) of the plot, and
the imaginary parts in the bottom half of the plot. The basis functions with
$m=0$ are wholly real. In a shapelet decomposition, all of the basis functions
are weighted by a complex number, whose magnitude determines the strength of a
component and whose phase sets its orientation. Arrows indicate the
``bleeding'' of power into four adjacent shapelet coefficients when a small
shear is applied. \textit{Right panel:} Reconstruction of irregular HDF
galaxies. Accurate models can be produced for even these peculiar shapes, using
$n_{\rm max}$ between 12 and 15, to leave image residuals entirely consistent
with noise.} \label{fig:basis} \end{figure}

The shapelets technique is based around the decomposition of a galaxy image
$f(r,\theta)$ into a weighted sum of (complete) orthogonal basis functions
\begin{equation} \label{eqn:decomp}
f(r,\theta) = \sum_{n=0}^\infty \sum_{m=-n}^{n} f_{n,m}
\chi_{n,m}(r,\theta;\beta) ~, \nonumber
\end{equation}

\noindent where $f_{n,m}$ are the ``shapelet coefficients''. The polar shapelet
basis functions $\chi_{n,m}(r,\theta)$ are shown in figure~\ref{fig:basis}.
These are successive perturbations around a Gaussian of width $\beta$
(equivalent to $r_g$ in KSB), parameterised by indices $n$ and $m$. The
mathematics of a shapelets is somewhat analogous to Fourier synthesis, but with
a compact support well-suited to the modelling of localised galaxies. For
example, a shapelet decomposition can similarly be truncated to eliminate the
highly oscillatory basis functions that correspond noise in the original image.

Note that figure~\ref{fig:basis} takes a convenient shorthand form. The basis
functions are only defined if $n$ and $m$ are both even or both odd but, for
clarity, their images have been enlarged into the spare adjacent space. The
polar shapelet basis functions and coefficients are also complex numbers.
However, the constraint that a combined image should be a wholly real function
introduces degeneracies: $f_{n,m}=f^*_{n,-m}$ and the coefficients with $m=0$
are wholly real. The top half of figure~\ref{fig:basis} with $m\geq 0$ shows
the real parts of the basis functions; the bottom half $m<0$ shows the complex
parts.

Before finding the shapelet coefficients for an image, it is necessary to
specify the centre of the shapelet basis functions, and their scale size
$\beta$. Since shapelets form a complete basis, decomposition is \textit{
possible} at any value of $\beta$. However, there is definitely a preferred
scale for most galaxies, with which a faithful model can be produced using only
a small number of shapelet coefficients. \cite{shapelets3} have written an
algorithm to automatically decompose an arbitrary image into shapelets by
exploring values of $\beta$. It seeks a model of the image that leaves a
residual consistent with noise, and chooses the scale size that achieves that
goal using the fewest coefficients. The optimal centre of the basis functions
can be found simultaneously, by shifting the basis functions so that the
model's unweighted centroid is zero. The procedure can also deal analytically
with the pixellisation of observational data, and perform
deconvolution from a Point-Spread Function. Its success at faithfully modelling
of even irregular HDF galaxies is demonstrated in the right-hand panel of
figure~\ref{fig:basis}. A complete IDL software package to implement the
shapelets decomposition of arbitrary images, and to perform analysis and
manipulation in shapelet space, can be downloaded from {\tt
www.astro.caltech.edu/$\sim$rjm/shapelets/}.

\firstsection\section{Why shapelets?}

It would be, of course, possible to analyse images using a more physically
motivated basis set, or traditional Sersic and Moffat radial profiles. However,
the shapelet basis functions are specifically chosen to simplify image analysis
and manipulation as encountered in weak gravitational lensing. As shown by
\cite{shapelets3}, shears, magnifications and convolutions are elegantly
represented in shapelet space as the mixture of power between an (almost)
minimal number of adjacent basis states. Shapelets are not motivated by their
{\it physics} but rather their {\it mathematics}. The burden of proof for
shapelets therefore shifts to the question of whether the central cusps
and extended wings of real galaxies can be faithfully modelled by a set of
functions based around a Gaussian. In fact, the recovery of galaxies' extended
wings is surprisingly complete with this algorithm. The process is helped by
the fact that the smooth shapelets basis functions can find faint but coherent
signal spread over many pixels, even though it may be beneath the noise level
in any given pixel (and therefore not detected by SExtractor).

\firstsection\section{Interpreting a polar shapelet decomposition}

A polar shapelet decomposition conveniently separates components of an image
that are intuitively different. The index $n$ describes the total number of
oscillations (spatial frequency) and also the size (radius) of the basis
function. The index $m$ describes the degree of rotational symmetry of the
basis functions.

Basis functions with $m=0$ are rotationally invariant. A circularly symmetric
object contains power only in these states; its flux and radial profile are
defined by the realtive values of its $m=0$ coefficients. An object containing
only $m=0$ states will be a useful place to start for lensing analysis because,
if galaxies' intrinsic ellipticities are uncorrelated, the ensemble average of
an unlensed population will indeed  be circularly symmetric. Even in a typical
galaxy, most of the power compactly occupies shapelet coefficients with low
$m$, and particularly those with $|m|=0$ or 2.

Basis functions with $|m|=1$ are invariant only under rotations of $360^\circ$.
These coefficients encode an object's centroid: their real and imaginary parts
correspond to displacements in the $x$ and $y$ directions. Alternatively, their
moduli correspond to an absolute distance, and their phases indicate a
direction.

Basis functions with $|m|=2$ are invariant under rotations of $180^\circ$, and
become negative versions of themselves under rotations of $90^\circ$. These are
precisely the properties of an ellipse. Indeed, an object's Gaussian-weighted
ellipticity $e\equiv e_1+ie_2$ is simply given by $f_{2,\pm2}$. Its unweighted
ellipticity is a combination of all of the $|m|=2$ shapelet coefficients.
Ellipticity estimators can also incorporate coefficients with
$|m|=6,10,14,...$, because their basis functions also contain at least the
necessary symmetries.

\firstsection\section{The effect of weak gravitational lensing in shapelet space}

All linear transformations can be described in shapelet space by the mixing of
power between a few adjacent shapelet coefficients. For example, let us begin
with an object containing power in just its $f_{4,0}$ coefficient. This is
indicated by the arrows overlaid on figure~\ref{fig:basis}. As the object is
sheared by $\gamma\equiv\gamma_1+i\gamma_2$, this power ``bleeds'' into four
nearby coefficients by an amount proportional to $\gamma$. Thus
$f_{2,2}~\rightarrow~f_{2,2}~+~\gamma~\times~\mathrm{constant}~\times~f_{4,0}$
(recall that $f_{2,2}$ is complex). To first order, $f_{4,0}$ is unchanged. The
diagonal pattern of the arrows is identical across the $n$ {\textit vs} $m$
plane, although the constant varies as a function of $n$ and $m$. For more
details, see \cite{shapelets3}.

An initially circular object may contain power in all of its $m=0$
coefficients. After a small shear, it also contains power in its $|m|=2$
coefficients: the combination of circularly-symmetric plus quadrupole states
produces an ellipse. Weak shear estimators primarily involve combinations of
the $|m|=2$ shapelet coefficients. For example, the $f_{2,2}$ coefficient is
the KSB ellipticity estimator. For a circularly symmetric object, this will
have been affected under the shear by the initial values of $f_{0,0}$ and
$f_{4,0}$. A weighted combination of these two (real) coefficients gives the
trace of the KSB $P^\gamma$ shear susceptibility tensor (ignoring terms
involving correction for PSF anisotropy).

A non-circularly symmetric object can also contain nonzero $|m|=4$
coefficients. Under a shear, the $f_{4,\pm 4}$ coefficients affect the
$f_{2,2}$ coefficient (plus some $|m|=6$ coefficients) to order $\gamma^*$.
Indeed, $f_{4,\pm 4}$ are the off-diagonal components of the KSB shear
susceptibility tensor. Unfortunately, the complex conjugation of $\gamma$ mixes
the $\gamma_1$ and $\gamma_2$ signals between the real and imaginary parts of
the $f_{2,2}$. It becomes impossible to disentangle the two components of
shear; and KSB can only work by averaging the shapes of many galaxies, to
ensure that the population's initial $|m|=4$ coefficients are precisely zero.

Using shapelets, \textit{every} shapelet coefficient with $|m|=2$ can provide a
statistically independent ellipticity estimator. Each of these has an effective
$P^\gamma$ involving its adjacent $m=0$ and $|m|=4$ coefficients. Multiple
shear estimators are very useful. Firstly, they can act as a consistency check
to examine measurement errors within each object. They can also be combined to
increase S/N: either by a simple average, or in more sophisticated ways that 
remove some of the biases of KSB (while staying stay linear in flux). For
example, it is possible to take a linear combination of $|m|=2$ coefficents
that is independent of the choice of $\beta$. However, the most successful
estimator involves a ``multiple multipole'' combination of $|m|=2,6,10,\dots$
shapelet coefficients that has $P^\gamma={\rm flux}$. This is an exciting
result for weak lensing, solving all of the problems with KSB's $P^\gamma$
listed in \S\ref{intro}. An object's flux is its zeroth-order moment, which can
be measured with less noise; it is a single, real number; and this shear
susceptibility is unchanged by a (pure) shear. We can therefore form shear
estimators using individual galaxies rather than having to average over a
population ensemble. The method works stably for any galaxy morphology, because
it does not rely on objects' initially having zero $|m|=4$ coefficients. This is
particularly important when the calibration is perfomed on simple image
simulations using elliptical galaxies with concentric isophotes.

As a final note of caution, gravitational lensing does not apply a pure shear:
it also applies a magnification, of the same order as $\gamma$. The enlargement
caused by a lensing magnification mixes power between a small number of nearby
shapelet coefficients. However, an enlargement is also equivalent to a increase
of $\beta$: this effect is therefore eliminated from lensing measurements using
a decomposition method with an adaptative choice of $\beta$. KSB and shapelet
shear measurements are also insensitive to the flux amplification, because
they are all formed from one linear sum of coefficients divided by another.

\firstsection\section{Results}

\begin{figure}
\begin{center}
\includegraphics[width=43mm]{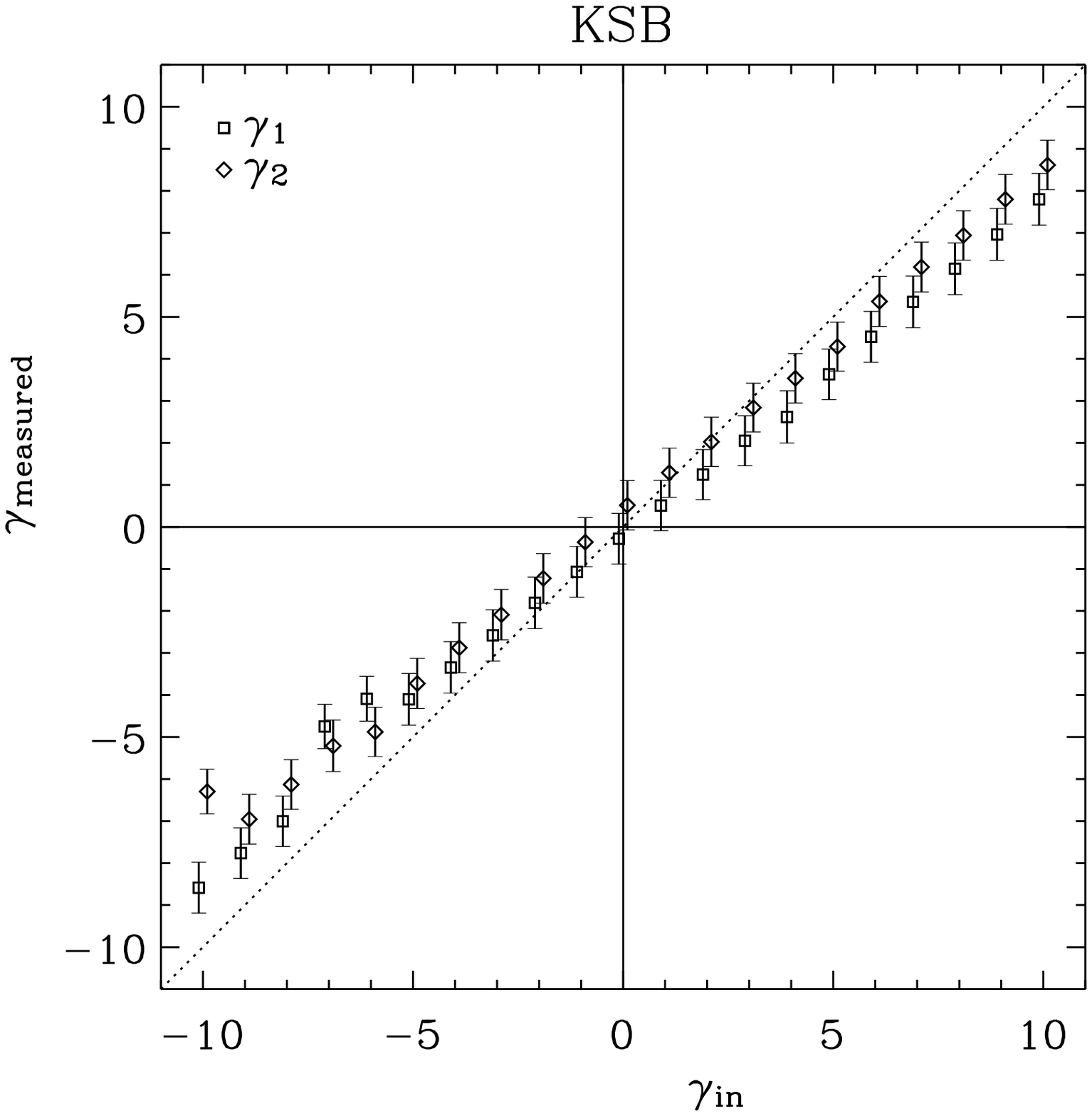}
\includegraphics[width=43mm]{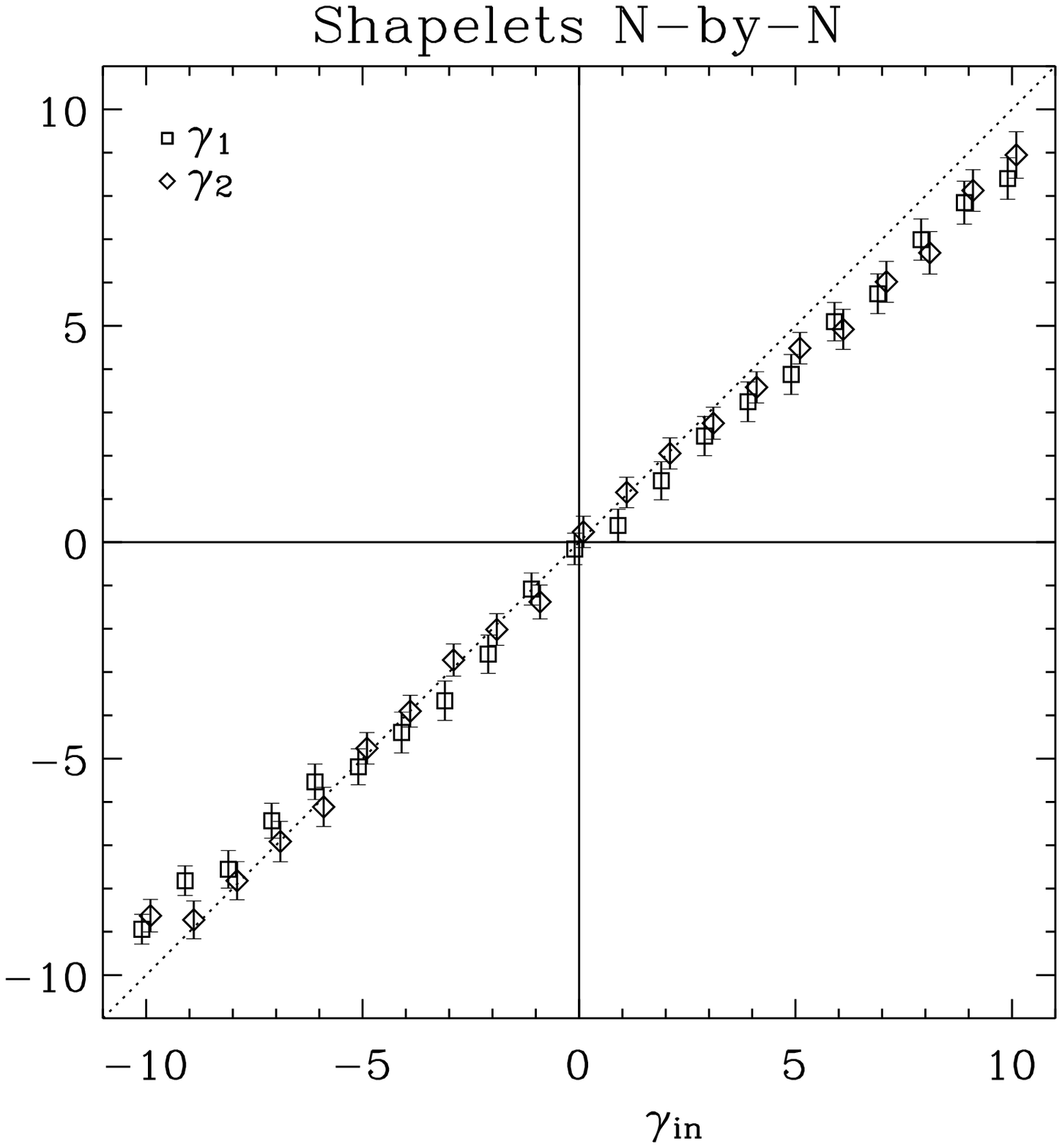}
\includegraphics[width=43mm]{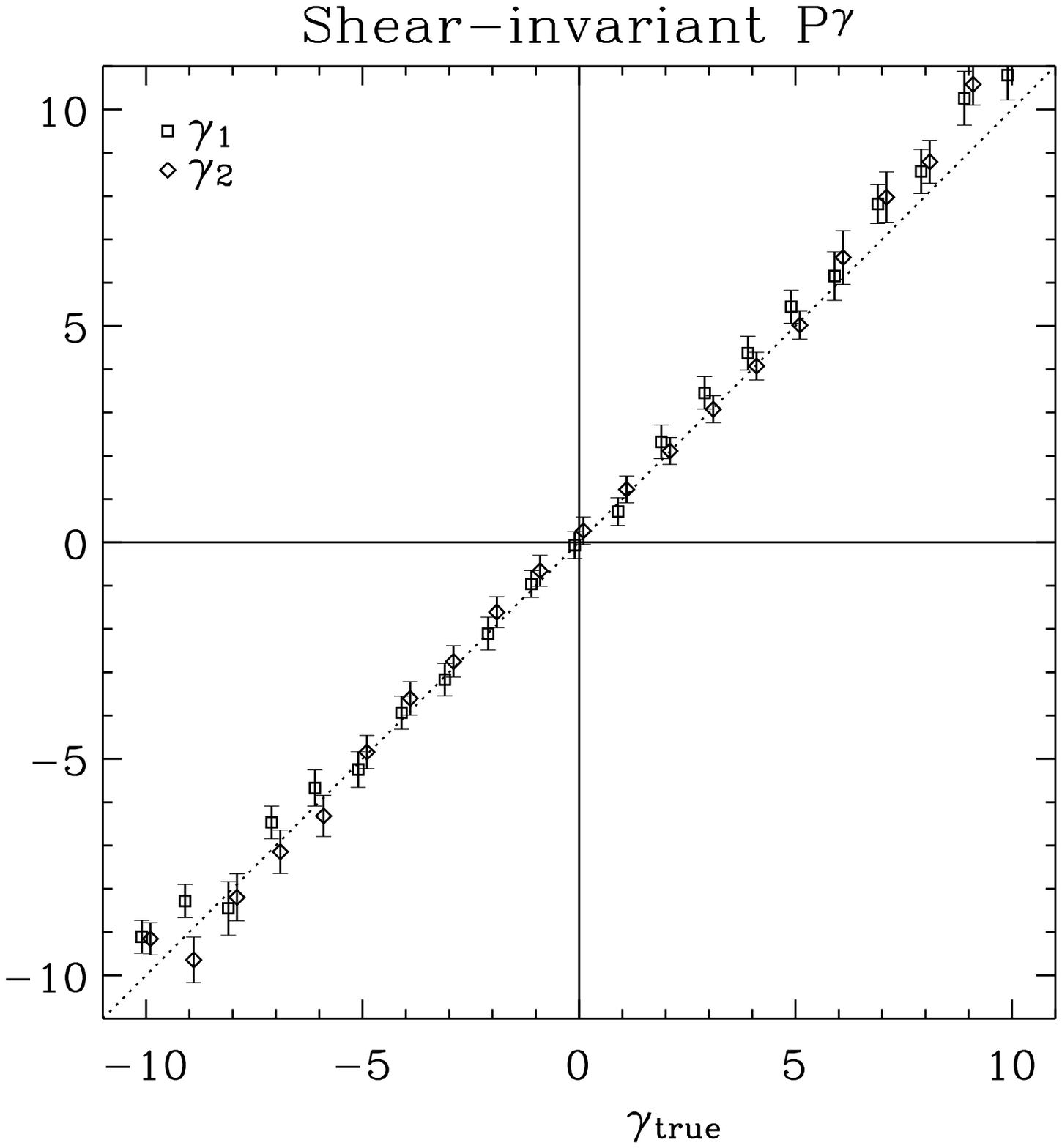}
\end{center}

\caption{\textit{Left panel:} recovery of a known shear signal using KSB.
Errors are at the $\sim10$\% level. The calibration bias has to be measured
using simulated images, then corrected for in real data. \textit{Middle panel:}
shear recovery by combining shapelet coefficients to create multiple, KSB-like
estimators. The S/N improves, but the calibration bias remains. \textit{Right
panel:} shear recovery using a more sophisticated, ``multiple multipole''
shapelets-based shear estimator. This is precise enough to detect deviations
from the weak shear approximation at high values of $|\gamma_{\rm in}|$.}

\label{fig:results}
\end{figure}

To test (and calibrate) various shear measurement methods, we have created
simulated images containing a known shear signal, $\gamma_{\rm in}$. We can
then compare measured values to the true value. Our simulated images mimic the
depth, pixellisation and PSF of the HDF, but the galaxies are simply
parameterised by concentric ellipses with an exponential radial profile. Such
objects are chosen to make the test especially challenging for shapelets-based
methods: their central cusps and extended wings of such objects will be hard to
match, while the concentric isphotes improve the prospects for KSB, that
effictively measures shear at only one fixed radius. We have created many
7.5~square degrees simulated images, each containing a constant input signal in
one component of shear, and zero in the other. Every shear measurement (and
each point in figure~\ref{fig:results}) can therefore be performed as an
average over a realistic population of galaxy sizes, magnitudes and intrinsic
ellipticities.

The shear measured by a real KSB pipeline, $\gamma_{\rm measured}$, is shown in
the left-hand panel of figure~\ref{fig:results}. Almost identical results can
be reproduced using the shapelets software to imitate KSB. The statistical
errors are quite large and there is calibration bias, as already noticed by
\cite{baconsims}. The value of the calibration factor can vary as a function of
exposure time and galaxy morphology, and therefore needs to be calibrated using
realistic simulated images. The precision of the KSB method is therefore
dependent upon the fidelity of the simulated images used to test it.

Using the shapelets formalism, we can derive many statistically independent
shear estimators for each object. The middle panel of figure~\ref{fig:results}
shows their (noise-weighted) average. This does indeed have higher S/N than the
KSB measurement; however, it still exhibits the familiar calibration bias. The
right-hand panel of figure~\ref{fig:results} shows results for the multiple
multipole shear estimator. This is very sensitive to weak shears: indeed, a
measurement of the components of shear that are not shown in
figure~\ref{fig:results} (which are all zero) gives 0.06\%$\pm$0.10\%. For
large input shear values, $|\gamma_{\rm in}|>6$\%, the precision of this
shapelets-based shear estimator is also sufficient to detect deviations from
the weak shear approximation.

\firstsection\section{Conclusions}

A shapelet decomposition parameterises {\it all} of an object's shape
information, in a convenient and intuitive form. Several shear estimators can
be formed from combinations of shapelet coefficients. These are not only more
accurate than KSB, but also more stable. In particular, the use of higher order
moments to analytically remove any calibration factor reduces the reliance of
older methods upon simulated images to faithfully model all aspects of
observational data.

\begin{acknowledgments}

The authors are pleased to thank Joel Berg\'{e}, Alain Boinissent, Tzu-Ching
Chang, Dave Goldberg, Will High, Konrad Kuijken and Molly Peeples for helpful
discussions.

\end{acknowledgments}

\firstsection

\end{document}